\documentstyle[12pt,a4,psfig]{article}
\begin{document}
\baselineskip 6.5mm

\begin{center}
\large \bf Duality and Enhanced Gauge Symmetry in 2+1 Dimensions
\end{center}

\vspace{5mm}

\begin{center}
Taichi Itoh$^{*}$\footnote{Email address: taichi@knu.ac.kr},
Phillial Oh$^\dagger$\footnote{Corresponding author. Email address: 
ploh@dirac.skku.ac.kr},
and Cheol Ryou$^\dagger$\footnote{Email address: cheol@newton.skku.ac.kr}
\end{center}

\begin{center} 
\it $^*$Department of Physics, Kyungpook National University, 
Taegu 702-701, Korea\\
$^\dagger$Department of Physics and Institute of Basic Science, 
Sungkyunkwan University, Suwon 440-746, Korea
\end{center}

\vspace{5mm}

\begin{center}
Abstract
\end{center}

We investigate the enlarged CP(N) model in 2+1 dimensions. This is a hybrid of 
two CP(N) models coupled with each other in a dual symmetric fashion, and it 
exhibits the gauge symmetry enhancement and radiative induction of the finite 
off-diagonal gauge boson mass as in the 1+1 dimensional case. We solve the 
mass gap equations and study the fixed point structure in the large-N limit. 
We find an interacting ultraviolet fixed point which is in contrast with the 
1+1 dimensional case. We also compute the large-N effective gauge action 
explicitly.

\vspace{10mm}

\vspace{30mm}
 \noindent{SKKUPT-04/2001, April 2001}

\newpage

\renewcommand{\theequation}{\arabic{section}\mbox{.}\arabic{equation}}

\section{Introduction}

The nonlinear sigma models have proved to be a very useful theoretical 
laboratory to study many important asymptotsubjects such as spontaneous 
symmetry breaking \cite{call, band}, asymptotic freedom and instantons in QCD
\cite{wilz,bela,cole}, the dynamical generation of gauge bosons
\cite{dada}, target space duality in string theory \cite{gree,give}, and many
others \cite{zakr}. Recently, some new properties have been explored in 
relation with the dynamical generation of gauge bosons, that is, the gauge 
symmetry enhancement and radiatively induced finite gauge boson mass in 1+1 
dimensions \cite{ior}. It is well-known that the $CP(N)\equiv SU(N)/SU(N-1) 
\times U(1)$ model \cite{golo} is the prototype of nonlinear sigma model with 
dynamical generation in which the  auxiliary $U(1)$ gauge field becomes 
dynamical through the radiative corrections in the large-$N$ limit 
\cite{dada}. In the recently proposed extension \cite{ior} of the $CP(N)$ 
model, two 
complex projective spaces with different coupling constants have mutual 
interactions  which are devised in such a way to preserve the duality between 
the two spaces. In addition to the two auxiliary $U(1)$ gauge fields which 
stand for each complex projective space, one extra auxiliary complex gauge 
field is introduced to derive the interactions with duality. It turns out that 
when the two coupling constants are equal, the extended model becomes the 
nonlinear sigma model with the target space of Grassmann manifold 
$Gr(N,2)=SU(N)/SU(N-2)\times U(2)$ \cite{gava}. 

It was shown in Ref. \cite{ior} that together with the two auxiliary $U(1)$ 
gauge fields this complex field becomes dynamical through radiative 
corrections. Moreover, in the self-dual limit where the two
running coupling constants become equal,  they become massless and combine 
with the two $U(1)$ fields to yield the $U(2)$ Yang-Mills theory. That is, the 
gauge symmetry enhancement has occurred in the self-dual limit. Away from this 
limit, the complex gauge field becomes massive. It was noted that this mass is 
radiatively induced through the loop corrections, and it assumes a finite 
value which is independent of the regularization scheme employed. 
This could provide an  alternative approach of providing the gauge boson
mass to the conventional Higgs mechanism. Therefore, it is
important to attempt to extend the previous 1+1 dimensions results of Ref. 
\cite{ior} in order to check whether this is also viable in  various other 
dimensions. In this paper, we take a first step, 
and extend the previous analysis to 2+1 dimensions.
Even though the 3+1 dimensional analysis awaits for some realistic 
applications, it has to be recalled that the $CP(N)$ model in 2+1
dimensions \cite{rose} has many extra interesting  properties such as
non-perturbative renormalizability despite of appearance of linear divergence, 
a non-trivial UV fixed point and second order phase transition \cite{aref}, 
and the induction of the Maxwell-Chern-Simons theory through the higher 
derivative interactions of renormalizable Wess-Zumino-Witten model 
\cite{itoh}. Therefore, the analysis carried out in this paper is expected to
shed light on the new aspects of 2+1 dimensional nonlinear sigma model in its 
own right.

The content of the paper is organized as follows. In Section 2, we review the
classical feature of the coupled dual $CP(N)$ model, and elaborate on the 
model in terms of coadjoint orbit approach. In Section 3, we solve 
the large $N$ mass gap equations, and find that there exist four phases of 
second order phase transition which are separated by UV fixed lines.
In Section 4, we discuss large $N$ renormalization and fixed point structure
of the vacua. In Section 5, we carry out the path integration
explicitly, and compute the $U(1) \times U(1)$ gauge invariant effective 
action in the unbroken phase. We show that the two point vacuum polarization 
graphs yield finite mass terms for the gauge fields which vanish at the 
self-dual limit, and the gauge symmetry is enhanced to $U(2)$ symmetry.
Section 6 includes conclusion and discussion. The dimensional regularization 
of vacuum polarization function is presented in Appendix A. We will show the 
detail of three- and four-point gauge vertices in Appendix B in the space-time 
dimensionality $2\le D\le 4$.

\section{Model and symmetry}
\setcounter{equation}{0}

We start from the Lagrangian written in terms of the $N\times2$ matrix $Z$
such that \cite{ior}
\begin{equation}
{\cal L}=\frac{1}{g^2}{\rm tr}
\left[(D_\mu Z)^\dagger (D^\mu Z)-\lambda(Z^\dagger Z-R)\right],
\label{lag1}
\end{equation}
where $\lambda$ is a $2\times2$ hermitian matrix which transforms as an 
adjoint representation under the local $U(2)$ transformation. 
The $R$ is a $2\times2$ matrix given by
\begin{equation}
R=\left[\,{r \atop 0}\quad{0 \atop r^{-1}}\right],
\end{equation}
with a real positive $r$.
The covariant derivative is defined consistently as 
$D_\mu Z \equiv\partial_\mu Z -Z\tilde{A}_\mu$ with a $2\times2$ 
anti-hermitian matrix gauge potential 
$\tilde{A}_\mu\equiv-i\tilde{A}_\mu^a T^a$ 
associated with the local $U(2)$ symmetry. 
We assign each components of $\lambda$ and $\tilde{A}_\mu$ as follows.
\begin{equation}
\lambda=\left[\,{\lambda_1 \atop \lambda_3^*}\quad
{\lambda_3 \atop \lambda_2}\right],\quad
\tilde{A}_\mu =-i\left[\,{A_\mu \atop \frac{1}{2}C_\mu^*}\quad
{\frac{1}{2}C_\mu \atop B_\mu}\right]. \label{gcom}
\end{equation}
The $Z$ field is made from two complex $N$-vectors $\psi_1$ and $\psi_2$ 
such that 
\begin{equation}
Z=\left[\psi_1, \psi_2\right],\quad\longleftrightarrow\quad 
Z^\dagger=\left[{\psi_1^\dagger \atop \psi_2^\dagger}\right].
\end{equation}

The kinetic term of the Lagrangian (\ref{lag1}) is invariant 
under the local $U(2)$ transformation, while the $R$ with $r\neq1$ explicitly 
breaks the $U(2)$ gauge symmetry down to $U(1)_A \times U(1)_B$ 
where $U(1)_A$ and $U(1)_B$ are generated by $T^0 \pm T^3$, 
respectively. Thus the symmetry of our model is 
$[SU(N)]_{\rm global} \times [U(2)]_{\rm local}$ for $r=1$, 
while $[SU(N)]_{\rm global} \times 
[U(1)_A \times U(1)_B]_{\rm local}$ for $r\neq 1$.
The local symmetry group $H$ is $U(2)$ when $r=1$, 
and $U(1)_A \times U(1)_B$ when $r \neq 1$.
To see the geometry of target space, we rewrite the Lagrangian (\ref{lag1})
in terms of two coupling constant $g_1$ and $g_2$ defined by
$g\equiv\sqrt{g_1g_2}$ and $r\equiv g_2 /g_1$. 
Using the on-shell constraint $Z^\dagger Z=R$ and rescaling the fields by
\begin{eqnarray}
\frac{\psi_1}{g}\rightarrow
\frac{\psi_1}{g_1},~\frac{\psi_2}{g}\rightarrow \frac{\psi_2}{g_2},
~ \frac{C_\mu}{g}\rightarrow C_\mu,~\frac{C_\mu^*}{g}\rightarrow 
C_\mu^*, 
\end{eqnarray}
the Lagrangian (\ref{lag1}) can be rewritten as 
\begin{eqnarray}
{\cal L} &=& \frac{1}{g_1^2}
|(\partial_\mu +iA_\mu)\psi_1|^2 +\frac{1}{g_2^2}|(\partial_\mu
+iB_\mu)\psi_2|^2 +\frac{1}{4}\left(\frac{g_1}{g_2}+\frac{g_2}{g_1}\right)
C_\mu^* C^\mu \nonumber \\ &&
-i\frac{1}{\sqrt{g_1g_2}}C_\mu^* \psi_1^\dagger \partial^\mu \psi_2
-i\frac{1}{\sqrt{g_1g_2}}C_\mu \psi_2^\dagger \partial^\mu \psi_1 
-\frac{\lambda_3^*}{ g_1g_2}\psi_1^\dagger \psi_2 -\frac{\lambda_3}
{g_1g_2}\psi_2^\dagger \psi_1 \label{lag2} \\ &&
-\frac{\lambda_1}{g_1^2} (\psi_1^\dagger \psi_1 -1)
-\frac{\lambda_2}{g_2^2} (\psi_2^\dagger \psi_2 -1). \nonumber
\end{eqnarray}
The above Lagrangian describes two $CP(N)$ models 
each described by $\psi_1$, $g_1$ and $\psi_2, g_2$ 
coupled through the derivative coupling.
There is a manifest dual symmetry between sectors 1 and 2,
$A_\mu$ and $B_\mu$, $C_\mu$ and $C^*_\mu$, and 
$\lambda_3$ and $\lambda_3^*$.
Eliminating the auxiliary fields through the equations of motion, and
substituting back into the Lagrangian, we obtain modulo the on-shell
constraints
\begin{eqnarray}
{\cal L}^\prime=\sum_{i=1}^2\frac{1}{g_i^2} \left[|\partial_\mu\psi _i|^2
+(\psi _i^\dagger\partial_\mu\psi _i)(\psi _i^\dagger\partial_\mu\psi _i)
\right]+\frac{2}{q}\sum_{i,j=1}^2{^\prime}\frac{1}{g_ig_j}
(\psi _i^\dagger\partial_\mu\psi _j)(\psi _j^\dagger\partial_\mu\psi _i),
\label{lagbq}
\end{eqnarray}
where $q=g_2/g_1~+ g_1/g_2$ and the prime in the third sum denotes that the 
sum is restricted to $i\neq j$ indices.
We notice that the target space geometry of the Lagrangian (\ref{lagbq})
with $q=2$ can be understood in the coadjoint orbit approach
\cite{oh1,bal} to 
nonlinear sigma model. In terms of coadjoint orbit variables 
\begin{eqnarray}
Q=\frac{1}{i}\sum_{i=1}^2\frac{1}{g_i}\left(\psi_i\psi_i^\dagger
-\frac{1}{N}I\right),~~~~\psi_i^\dagger\psi_j=\delta_{ij},
\end{eqnarray}
the Lagrangian (\ref{lagbq}) with $q=2$ can be rewritten as
\begin{eqnarray}
{\cal L}_Q= -\frac{1}{2}\mbox{tr}(\partial_\mu Q)^2.
\label{deform}
\end{eqnarray}
In the above Lagrangian (\ref{deform}), the equal coupling $g_1= g_2$ 
corresponds the target space of Grassmann manifold
$Gr(N,2)$, whereas the non-equal couplings $g_1\neq g_2$ 
to the flag manifold ${\cal M}=SU(N)/SU(N-2)\times U(1)\times U(1)$
\cite{helg}. Therefore, the generic $q\neq 2$ case of the Lagrangian
(\ref{lagbq}) is a deformation of the flag manifold model.

In order to carry out the path integration in the large $N$ limit, we rewrite
the Lagrangian (\ref{lag1}) in terms of a $2\times 2$ hermitian matrix such
that 
\begin{equation}
{\cal L}=\frac{1}{g^2}
\,[\psi_1^\dagger,\psi_2^\dagger]\,(M^T\otimes I)
\left[{\psi_1 \atop \psi_2}\right]
+\frac{r}{g^2}\lambda_1 +\frac{1}{rg^2}\lambda_2,
\label{lag3}
\end{equation}
where $\langle\cdots\rangle$ denotes a trace of an $N\times N$ matrix. 
The $2\times 2$ matrix operator $M$ is given by
\begin{eqnarray}
M &\equiv& G^{-1}-\Gamma(\tilde{A}), \label{matrix}\\
G^{-1} &\equiv& -\Box -\lambda\,\,=\,\,
\left[\,{-\Box -\lambda_1 \atop -\lambda_3^*}\quad
{-\lambda_3 \atop -\Box -\lambda_2}\right], \label{Ginv}\\
\Gamma(\tilde{A}) &\equiv& 
-\tilde{A}_\mu \hat{\partial}^\mu
+\tilde{A}_\mu \tilde{A}^\mu. \label{gamma}
\end{eqnarray}
where the differential operator 
$\hat{\partial}^\mu\equiv\partial^\mu -\overleftarrow{\partial^\mu}$ 
must be regarded as not operating on the gauge potential $\tilde{A}_\mu$. 
In terms of $A_\mu$, $B_\mu$ and $C_\mu$ fields, 
all components of the matrix $M$ are written as
\begin{eqnarray}
M_{11} &=& -\partial^2 -\lambda_1 
-iA_\mu \hat{\partial}^\mu
+A_\mu A^\mu +\frac{1}{4}C_\mu^* C^\mu, \label{m11}\\ 
M_{22} &=& -\partial^2 -\lambda_2 
-iB_\mu \hat{\partial}^\mu
+B_\mu B^\mu +\frac{1}{4}C_\mu^* C^\mu, \label{m22}\\
M_{12} &=& -\lambda_3 
-\frac{1}{2}iC_\mu \hat{\partial}^\mu
+\frac{1}{2}C_\mu (A^\mu +B^\mu), \label{m12}\\
M_{21} &=& -\lambda_3^* 
-\frac{1}{2}iC_\mu^* \hat{\partial}^\mu
+\frac{1}{2}C_\mu^* (A^\mu +B^\mu).\label{m21}
\end{eqnarray}
Here we have never used the on-shell constraint so that the quadratic 
term of $C_\mu^* C^\mu$ has been absorbed into the matrix $M$. 
The last terms in Eqs.\ (\ref{m12}) and (\ref{m21}) were missing in the 
Lagrangian (\ref{lag2}) due to the on-shell constraint but they are 
essential to recover the gauge invariance of the off-shell Lagrangian 
(\ref{lag3}). 
We use the off-shell Lagrangian (\ref{lag3}) in order to preserve the gauge
invariance in every step of computation.

\section{Large-N gap equations}
\setcounter{equation}{0}

The large $N$ effective action is given by path integrating $Z$ and 
$Z^\dagger$, or equivalently $\psi_1$, $\psi_1^\dagger$, $\psi_2$, and 
$\psi_2^\dagger$. We obtain
\begin{equation}
S_{\rm eff}=\int_x {\cal L}+iN\ln{\rm Det}M. \label{eff1}
\end{equation}
The global $U(N)$ symmetry enables us to choose the VEV vectors 
$\langle\psi_1\rangle$ and $\langle\psi_2\rangle$ to be real $N$-vectors 
and we can set all $\lambda_1$, $\lambda_2$, and $\lambda_3$ to be real 
without loss of generality. The large-$N$ effective action is determined as
\begin{equation}
V_{\rm eff}=-\frac{1}{N\Omega}S_{\rm eff}
[\psi_{1,2}=\vec{v}_{1,2},\lambda_{1,2,3}=m^2_{1,2,3},
\tilde{A}_\mu=0],
\end{equation}
where $\vec{v}_1$, $\vec{v}_2$ are real $N$-vectors and $\Omega$ denotes 
the space-time volume. We obtain
\begin{equation}
V_{\rm eff}=\frac{m^2_1}{Ng^2}(\vec{v}_1^2 -r)
+\frac{m^2_2}{Ng^2}(\vec{v}_2^2 -r^{-1})
+\frac{2m^2_3}{Ng^2}\vec{v}_1 \cdot \vec{v}_2
-i\Omega^{-1}\ln{\rm Det}\,G^{-1},
\end{equation}
The gap equations are schematically given as follows.
\begin{eqnarray}
\frac{\partial V_{\rm eff}}{\partial \vec{v}_1} &=& 
\frac{2}{Ng^2}(m^2_1 \vec{v}_1 +m^2_3 \vec{v}_2)=0,
\label{gap1}\\
\frac{\partial V_{\rm eff}}{\partial \vec{v}_2} &=& 
\frac{2}{Ng^2}(m^2_2 \vec{v}_2 +m^2_3 \vec{v}_1)=0,
\label{gap2}\\
\frac{\partial V_{\rm eff}}{\partial m^2_3} &=& 
\frac{2}{Ng^2}\vec{v}_1 \cdot \vec{v}_2 
-\int\! \frac{d^D k}{(2\pi)^D}
\frac{2m^2_3}{(k^2 +m^2_+)(k^2 +m^2_-)}=0,
\label{gap3}\\
\frac{\partial V_{\rm eff}}{\partial m^2_1} &=& 
\frac{1}{Ng^2}(\vec{v}_1^2 -r)+\int\! \frac{d^D k}{(2\pi)^D}
\frac{k^2 +m^2_2}{(k^2 +m^2_+)(k^2 +m^2_-)}=0,
\label{gap4}\\
\frac{\partial V_{\rm eff}}{\partial m^2_2} &=& 
\frac{1}{Ng^2}(\vec{v}_2^2 -r^{-1})+\int\! \frac{d^D k}{(2\pi)^D}
\frac{k^2 +m^2_1}{(k^2 +m^2_+)(k^2 +m^2_-)}=0
\label{gap5},
\end{eqnarray}
where the loop momenta are euclideanized
and $m^2_\pm$ are given in terms of $m^2_{1,2,3}$ by
\begin{equation}
m^2_+ +m^2_- =m^2_1 +m^2_2,\quad
m^2_+ m^2_- =m^2_1 m^2_2 -m^4_3.
\label{lambda1}
\end{equation} 
We focus on $D=3$ case below. 

First we have to regularize the divergent integrals in the gap equations.
We separate out the ultraviolet divergence in (\ref{gap4}) such that
\begin{equation}
\int^\Lambda\! \frac{d^D k}{(2\pi)^3}\frac{1}{k^2}
+\int\! \frac{d^D k}{(2\pi)^3}
\left[\frac{k^2 +m^2_2}{(k^2 +m^2_+)(k^2 +m^2_-)}
-\frac{1}{k^2}\right],
\label{int1}
\end{equation}
which is calculated to be
\begin{equation}
\frac{1}{2\pi^2}\Lambda -\frac{1}{4\pi}
\frac{m^2_1 +m_+ m_-}{m_+ +m_-}.
\end{equation}
Then we obtain the properly regularized gap equations in $D=3$:
\begin{eqnarray}
m^2_1 \vec{v}_1 +m^2_3 \vec{v}_2 &=& 0,\label{gap6}\\
m^2_2 \vec{v}_2 +m^2_3 \vec{v}_1 &=& 0,\label{gap7}\\
\frac{1}{u}\Lambda\,\vec{v}_1 \cdot\vec{v}_2 &=& 
\frac{1}{4\pi}\frac{m^2_3}{m_+ +m_-},\label{gap8}\\
\frac{1}{u}\Lambda\,\vec{v}_1^2 &=&
\left(\frac{r}{u}-\frac{1}{u^*}\right)\Lambda
+\frac{1}{4\pi}\frac{m^2_1 +m_+ m_-}{m_+ +m_-},\label{gap9}\\
\frac{1}{u}\Lambda\,\vec{v}_2^2 &=&
\left(\frac{1}{ur}-\frac{1}{u^*}\right)\Lambda
+\frac{1}{4\pi}\frac{m^2_2 +m_+ m_-}{m_+ +m_-},\label{gap10}
\end{eqnarray}
where we have introduced the dimensionless coupling $u\equiv Ng^2 \Lambda$ 
and $u^* \equiv 2\pi^2$.

Suppose that $m_3 \neq0$ is a solution to the gap equations. 
The equations (\ref{gap6}) and (\ref{gap7}) yield that $\vec{v}_1$ and 
$\vec{v}_2$ are anti-parallel to each other, say specifically,
\begin{equation}
\vec{v}_2 =-\frac{m^2_1}{m^2_3}\vec{v}_1,\quad
\vec{v}_1 =-\frac{m^2_2}{m^2_3}\vec{v}_2,
\label{lambda2}
\end{equation}
of which iterative substitution provides $m^2_3 =m_1 m_2$
so that we can set $m_+ =\sqrt{m^2_1 +m^2_2}$, $m_- =0$. 
Substituting (\ref{lambda2}) into (\ref{gap8}), we obtain
\begin{equation}
\frac{1}{u}\Lambda\,\vec{v}_1^2 = -\frac{1}{4\pi}
\frac{m_2^2}{\sqrt{m^2_1 +m^2_2}}.
\end{equation}
The left hand side is positive, whereas the right hand side is 
negative. This result is obviously inconsistent in itself and we therefore 
conclude that $m_3 \neq0$ is not a solution to the gap equations.

Setting $m_3 =0$ in (\ref{lambda1}), we can choose for example 
$m_+ =m_1$, $m_- =m_2$. Then the gap equations are 
simplified such that
\begin{eqnarray}
m^2_1 \vec{v}_1 &=& 0,\label{gap11}\\
m^2_2 \vec{v}_2 &=& 0,\label{gap12}\\
\vec{v}_1 \cdot \vec{v}_2 &=& 0,\label{gap13}\\
\frac{1}{u}\Lambda\,\vec{v}_1^2 &=& 
\left(\frac{r}{u}-\frac{1}{u^*}\right)\Lambda
+\frac{1}{4\pi}m_1,\label{gap14}\\
\frac{1}{u}\Lambda\,\vec{v}_2^2 &=& 
\left(\frac{1}{ur}-\frac{1}{u^*}\right)\Lambda
+\frac{1}{4\pi}m_2\label{gap15}.
\end{eqnarray}
Note that $\vec{v}_1$ and $\vec{v}_2$ are perpendicular each other and 
may possibly break the global $SU(N)$ symmetry down to the 
$SU(N-2)\times U(1)\times U(1)$ symmetry.

\begin{figure}
\centerline{\hbox{\psfig{file=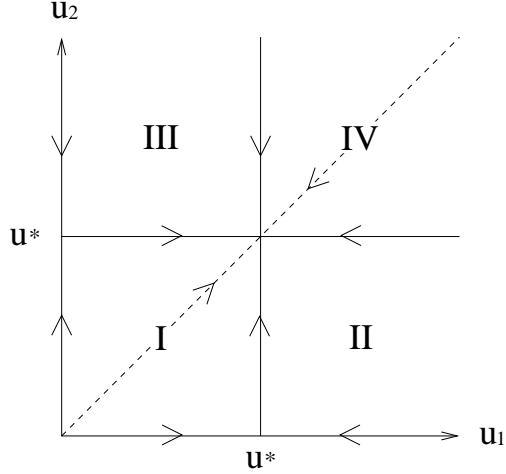,height=6.5cm}}}
\caption{Large-$N$ phase diagram in three-dimensions.}\label{fig1} 
\vspace{12pt}
\end{figure}

In order to simplify the following analysis, let us introduce $u_1\equiv u/r$ 
and $u_2\equiv ur$. The possible phases of vacuum are classified depending 
on the regions in the parameter space $(u_1,u_2)$ as follows 
(See Fig.~\ref{fig1}.).

\begin{list}{}{}
\item[I.] $u_1<u^*$ and $u_2<u^*$ $\,\longleftrightarrow\,$ 
$u/u^* <\min\{r,r^{-1}\}$

Since the right hand sides of both (\ref{gap14}) and (\ref{gap15}) become 
positive in this case, we have a solution: $\vec{v}_1 \neq\vec{0}$, 
$\vec{v}_2 \neq\vec{0}$ ($m_1 =0$, $m_2 =0$). The orthogonal 
condition (\ref{gap13}) tells us that this solution maximally breaks 
the global $SU(N)$ symmetry down to $SU(N-2)\times U(1)\times U(1)$.
From Eqs.\ (\ref{m11}), (\ref{m22}) we see that all gauge fields become 
massive due to $\langle\psi_1^\dagger \psi_1\rangle\neq0$ and 
$\langle\psi_2^\dagger \psi_2\rangle\neq0$ so that the gauge group $H$ 
is fully broken.

\item[II.] $u_1>u^*$ and $u_2<u^*$ $\,\longleftrightarrow\,$ 
$r< u/u^* <r^{-1}$

The right hand side of (\ref{gap14}) is not positive definite 
so that we have a solution: $\vec{v}_1 =\vec{0}$, $\vec{v}_2 \neq\vec{0}$ 
($m_1 \neq0$, $m_2 =0$). The $SU(N)$ symmetry is broken down 
to $SU(N-1)\times U(1)$. Eq.\ (\ref{m22}) tells us that 
$B_\mu$ and $C_\mu$ become massive due to 
$\langle\psi_2^\dagger \psi_2\rangle\neq0$, while $A_\mu$ remains massless 
as shown in Eq.\ (\ref{m11}). In terms of the $U(2)$ adjoint gauge fields, 
$A_\mu$ is written as $A_\mu =(\tilde{A}^{0}_\mu +
\tilde{A}^{3}_\mu)/2$ and is therefore regarded as a gauge filed 
associated with the $U(1)_A$ gauge symmetry. 
The local $H$ symmetry is broken down to the $U(1)_A$ symmetry.

\item[III.] $u_1<u^*$ and $u_2>u^*$ $\,\longleftrightarrow\,$ 
$r^{-1}< u/u^* <r$

As the case before we have a solution: $\vec{v}_1 \neq\vec{0}$, 
$\vec{v}_2 =\vec{0}$ ($m_1 =0$, $m_2 \neq0$). 
The $SU(N)$ symmetry is broken down to $SU(N-1)\times U(1)$.
Since $B_\mu$ only remains massless, the $H$ symmetry is broken 
down to the $U(1)_B$ symmetry.

\item[IV.] $u_1>u^*$ and $u_2>u^*$ $\,\longleftrightarrow\,$ 
$u/u^* >\max\{r,r^{-1}\}$

We only have a trivial solution: $\vec{v}_1 =\vec{0}$, 
$\vec{v}_2 =\vec{0}$ ($m_1 \neq0$, $m_2 \neq0$).
Both global $SU(N)$ and local $H$ symmetries remain unbroken. 

\end{list}

We notice that the four phases I,~II,~III,~IV are separated by the two 
critical lines $u_1 =u^*$ and $u_2 =u^*$ which arise as ultraviolet (UV)
fixed lines associated with the second order phase transitions 
after the large-$N$ renormalization of the effective potential.

\section{Renormalization and fixed point structure of the vacua}
\setcounter{equation}{0}

\begin{figure}
\centerline{\hbox{\psfig{file=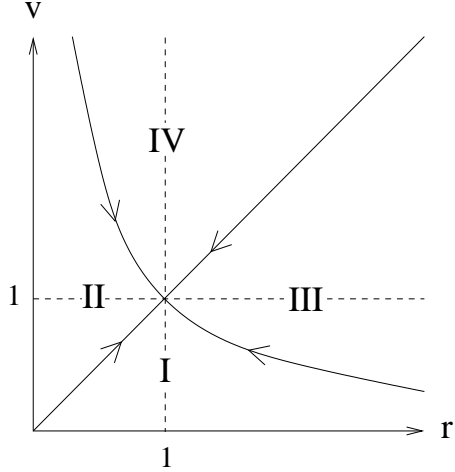,height=6.5cm}}}
\caption{Large-$N$ phase diagram in the $(r,v)$-plane.}\label{fig2} 
\vspace{12pt}
\end{figure}

The only UV divergences in the large-$N$ effective potential are those 
in the gap equations (\ref{gap14}), (\ref{gap15}) so that we 
impose the following renormalization conditions:
\begin{eqnarray}
\frac{d}{d\ln\Lambda}\left(\frac{1}{u_1}-\frac{1}{u^*}\right)\Lambda=0,
\label{rg1}\\
\frac{d}{d\ln\Lambda}\left(\frac{1}{u_2}-\frac{1}{u^*}\right)\Lambda=0.
\label{rg2}
\end{eqnarray}
This yields two decoupled renormalization group (RG) equations
\begin{eqnarray}
\frac{d u_1}{d\ln\Lambda} &=& u_1 \left(1-\frac{u_1}{u^*}\right),\\
\frac{d u_2}{d\ln\Lambda} &=& u_2 \left(1-\frac{u_2}{u^*}\right),
\end{eqnarray}
of which UV fixed points $u_1=u^*$ and $u_2=u^*$ can be identified with 
the two critical lines which separate the four different phases. 
Moreover, the intersection point
$(u_1,u_2)=(u^*,u^*)$ is 
conformally invariant. This situation is realized as a self-dual condition 
$r=1$ ($u_1 =u_2)$ which arises as a UV fixed line of the RG 
$\beta$-function for $r$. 
In terms of $v\equiv u/u^*$ and $r$, the RG 
equations are equivalently rewritten as two coupled equations:
\begin{eqnarray}
\frac{d v}{d\ln\Lambda} &=& v\left[1-\frac{1}{2}\left(r+\frac{1}{r}\right)
v\right],\label{rg3}\\
\frac{d r}{d\ln\Lambda} &=& \frac{1}{2}v\left(1-r^2\right),\label{rg4}
\end{eqnarray}
which show two relevant directions $v=r$ and $v=r^{-1}$ around 
$(r,v)=(1,1)$. In fact, if we substitute $v=r$ or $v=r^{-1}$ into 
(\ref{rg3}) and (\ref{rg4}), the two equations reduce to the equations
\begin{equation}
\frac{dr}{d\ln\Lambda}=\left\{
{\displaystyle \frac{1}{2}r(1-r^2)\quad\mbox{for}\quad v=r, \atop 
\displaystyle \frac{1}{2r}(1-r^2)\quad\mbox{for}\quad v=\frac{1}{r}.}
\right.
\end{equation}
This shows existence of the UV fixed point at $r=v=1$. The phase diagram in 
the $(r,v)$-plane is depicted in Fig.~\ref{fig2}.

\section{Large-N effective action and enhanced gauge symmetry}
\setcounter{equation}{0}

The large-$N$ effective action (\ref{eff1}) is schematically expanded 
such that
\begin{equation}
S_{\rm eff}=\int_x {\cal L}+iN\ln{\rm Det}\,G^{-1}
-iN\sum_{n=1}^{\infty}\frac{1}{n}{\rm Tr}\left[G\Gamma(\tilde{A})\right]^n.
\label{eff2}
\end{equation}
The boson propagator $G$ becomes a diagonal $2\times2$ matrix due to 
the gap equation solution $m_3 =0$. We neglect the fluctuation
fields coming from $\lambda_{1,2,3}$ around $m^2_{1,2,3}$ and 
consider the symmetric phase IV. 
In the following we study the diagrams up to four-point functions 
which are of the lowest order in the derivative expansion and cast into the 
Yang-Mills action of the enhanced $U(2)$ gauge symmetry at the 
self-dual limit r=1.

\subsection{Vacuum polarization and the off-diagonal gauge boson mass}

\begin{figure}
\centerline{\hbox{\psfig{file=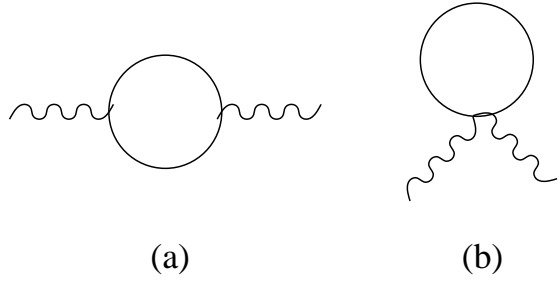,height=4cm}}}
\caption{Vacuum polarization diagrams. (a) $n=2$. (b) $n=1$.}\label{fig3} 
\vspace{12pt}
\end{figure}

We have two diagrams in Fig.~\ref{fig3}. 
They are combined into kinetic terms such that
\begin{eqnarray}
\mbox{(3a)}+\mbox{(3b)} &=& 
-iN\frac{1}{2}{\rm Tr}\left[G\tilde{A}_\mu \hat{\partial}^\mu
G\tilde{A}_\nu \hat{\partial}^\nu \right]
-iN{\rm Tr}\left[G\tilde{A}_\mu\tilde{A}^\mu \right]
\nonumber \\ &=& 
\frac{N}{2}\sum_{ij}\int_x\,\tilde{A}^\mu_{ij}(x)
\Pi_{\mu\nu}^{ij}(i\partial_x)\tilde{A}^\nu_{ji}(x),
\end{eqnarray}
where the vacuum polarization function $\Pi$ is given by
\begin{equation}
\Pi_{\mu\nu}^{ij}(p) = -\int\!\frac{d^3 k}{i(2\pi)^3}
\frac{(2k+p)_\mu (2k+p)_\nu}{(k^2 -m_i^2)[(k+p)^2 -m_j^2]}
+\int\!\frac{d^3 k}{i(2\pi)^3}\frac{2g_{\mu\nu}}{k^2 -m_i^2}.
\label{vacpol}
\end{equation}
which must be regularized so as to preserve the $U(1)_A \times U(1)_B$ 
gauge invariance which is manifest even when $r\neq 1$. 
The vacuum polarization function is calculated such that
\begin{equation}
\Pi_{\mu\nu}^{ij}(p)=
\left(g_{\mu\nu}-\frac{p_\mu p_\nu}{p^2}\right)\Pi_T^{ij}(p)
+\left(\frac{p_\mu p_\nu}{p^2}\right)\Pi_L^{ij}(p),
\label{mas}
\end{equation}
with the transverse function $\Pi_T$ and the longitudinal one $\Pi_L$ 
obtained as (See Appendix A)
\begin{eqnarray}
\Pi_T^{ij}(p) &\equiv& 
\frac{1}{2\pi}\left[\frac{m_i +m_j}{2}-\int_0^1 dx 
\sqrt{K}\right],
\label{mass}\\
\Pi_L^{ij}(p) &\equiv& \frac{\left(m_i^2 -m_j^2\right)^2}{8\pi p^2} 
\left[\frac{2}{m_i +m_j}-\int_0^1 dx 
\frac{1}{\sqrt{K}}\right],
\label{mass2}
\end{eqnarray}
where we have introduced $K \equiv x m_i^2 +(1-x) m_j^2 -x(1-x)p^2$. 
Each of $\Pi_T$ and $\Pi_L$ has a constant as the leading term 
in momentum expansion. Moreover we see that
\begin{eqnarray}
\Pi_T^{ij}(p) &=& c^{ij} + p^2 f_T^{ij} (p),\\
\Pi_L^{ij}(p) &=& c^{ij} + p^2 f_L^{ij} (p),
\end{eqnarray}
where the same constant $c^{ij}$ arises both in $\Pi_T$ and $\Pi_L$ 
and is determined as
\begin{equation}
c^{ij}=-\frac{(m_i -m_j)^2}{12\pi(m_i +m_j)}.
\end{equation}
Then the vacuum polarization can be written as
\begin{equation}
\Pi_{\mu\nu}^{ij}(p)=c^{ij} g_{\mu\nu} 
+(p^2 g_{\mu\nu}-p_\mu p_\nu) f_T^{ij}(p)
+p_\mu p_\nu f_L^{ij}(p),
\end{equation}
where both $c^{ij}$ and $f_L^{ij}$ vanish when $i=j$ 
so as to provide the $A$ ($B$) boson with the $U(1)_A$ ($U(1)_B$) 
gauge invariant kinetic term, while they remain nonzero when $i\neq j$ 
and provide the $C$ boson with the mass given by (See Appendix A)
\begin{equation}
M_C =\sqrt{\frac{-c^{12}}{f_T^{12}(0)}}
=\left|\frac{m_1^2 -m_2^2}{2}\right|\sqrt{\frac{5}
{m_1^2 +m_2^2 +3m_1 m_2}}.
\end{equation}
A couple of remarks are in order. Firstly, we note that
the above mass does not vanishes when $m_1\neq m_2$ which in turn implies
$r\neq 1$ from the mass gap equations (\ref{gap14}) and (\ref{gap15}).
It is also symmetric under the exchange of $m_1$ and $m_2$. 
At the self-dual limit $r=1$ ($m_1 =m_2 =m$), both $c^{12}$ and 
$\tilde{\Pi}_L^{12}$ become zero so that the off-diagonal $C$ boson 
becomes massless and combines into the enhanced $U(2)$ gauge fields 
together with the diagonal $A$, $B$ bosons.
Secondly, it should be emphasized that this mass generation of $C$ bosons is
a genuine quantum effect away from the self-dual line and the mass takes
a definite value in terms of the two mass scales without any ambiguity.
It is also independent of the regularization scheme employed. 
This unambiguity is in contrast with some other radiative corrections in 
quantum field theory which are finite but undetermined \cite{jack}.

The vacuum polarization diagrams in Fig.~\ref{fig3} finally provide 
the kinetic terms
\begin{eqnarray}
\mbox{(3a)}+\mbox{(3b)} &=& 
\frac{N}{4}\int_x~\Biggl[
-f_T^{11}(0) F_{\mu\nu} F^{\mu\nu}(A)
-f_T^{22}(0) F_{\mu\nu} F^{\mu\nu}(B)
-c^{12} C^\mu C_\mu^*
\nonumber \\
&&-\frac{1}{2} f_T^{12}(0)
\partial_{[\mu}C_{\nu ]}^* ~\partial^{[\mu} C^{\nu ]}
-f_L^{12}(0)\partial^\mu C^*_{\mu} \partial^\nu C_\nu
\Biggr]. \label{2point}
\end{eqnarray}
in the leading order of derivative expansion.
At the self-dual limit, $f_T^{11}(0)=f_T^{22}(0)=N/24\pi m$ and 
$c^{12}$, $f_L^{12}(0) \to 0$ so that the above kinetic terms are 
rearranged into
\begin{equation}
\mbox{(3a)}+\mbox{(3b)}=\frac{N}{96\pi m}\int_x\left[
{\rm tr}\left\{\partial_{[\mu}\tilde{A}_{\nu]}
\partial^{[\mu}\tilde{A}^{\nu]}\right\}
+O(\partial^2 \frac{\partial^2}{m^2})\right].
\end{equation}

\subsection{Three-point gauge vertices}

\begin{figure}
\centerline{\hbox{\psfig{file=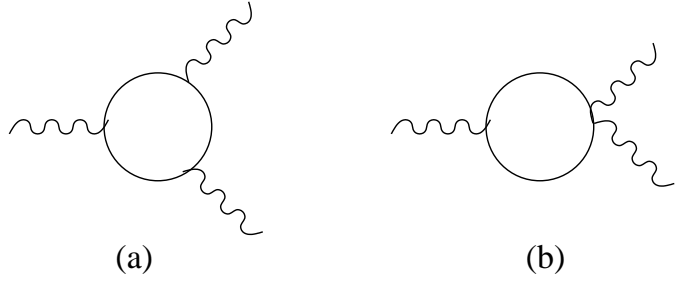,height=4cm}}}
\caption{Three-point gauge vertices. (a) $n=3$. (b) $n=2$.}\label{fig4} 
\vspace{12pt}
\end{figure}

Both of three-point gauge diagrams (a) and (b) in Fig.~\ref{fig4} 
contribute to the Yang-Mills action. 
They are given by the following integrals:
\begin{eqnarray}
\mbox{(4a)} &=& iN\frac{1}{3}{\rm Tr}\left[G\tilde{A}_\mu 
\hat{\partial}^\mu G\tilde{A}_\nu \hat{\partial}^\nu
G\tilde{A}_\rho \hat{\partial}^\rho \right], \label{4a}\\ 
\mbox{(4b)} &=& iN{\rm Tr}\left[G\tilde{A}_\mu \hat{\partial}^\mu
G\tilde{A}_\nu\tilde{A}^\nu\right]. \label{4b}
\end{eqnarray}
In the leading order of derivative expansion, 
they are calculated to be (See Appendix B)
\begin{eqnarray}
\mbox{(4a)}+\mbox{(4b)} &=& 
\frac{iN}{4}f_T^{12}(0) \int_x~\Biggl[ 
-\frac{1}{2}\Bigl(
W_{[\mu}C_{\nu ]}^* \partial^{[\mu}C^{\nu ]}
-\partial_{[\mu}C_{\nu ]}^* W^{[\mu}C^{\nu ]}
\Bigr)  \nonumber \\
&&-\frac{2}{3}(b_1 +b_2 \Bigr)
\Bigl(
W_{\mu}C^*_{\nu} \partial^{\nu}C^{\mu}
-\partial_{\mu}C^*_{\nu} W^{\nu}C^{\mu}
\Bigr)      \nonumber \\
&&-\frac{1}{3}\left(b_1 - 2b_2\right)
\Bigl(
W_{\mu}C^{*\mu} \partial_{\nu}C^{\nu} 
-\partial_{\mu}C^{*\mu} W_{\nu}C^{\nu}
\Bigr)   \nonumber \\
&&-\Bigl(1-\frac{b_1}{3}-\frac{b_2}{3}\Bigr)
C^\nu C^{*\mu}F_{\mu\nu}(A)
-\Bigl(1-\frac{b_1}{3}+\frac{b_2}{3}\Bigr)
C^{*\nu} C^{\mu}F_{\mu\nu}(B)
\Biggr], \label{3point}
\end{eqnarray}
where we have defined $b_1$, $b_2$ and $W_\mu$ such that
\begin{equation}
b_1 \equiv \frac{f_L^{12}(0)}{f_T^{12}(0)},\quad
b_2 \equiv \frac{M_C^2}{m_1^2 -m_2^2},\quad
W_\mu \equiv A_\mu-B_\mu.
\end{equation}
At the self-dual limit $r=1$ ($m_1 =m_2 =m$), Eq.\ (\ref{3point}) 
turns to the following simple form
\begin{equation}
\mbox{(4a)}+\mbox{(4b)}=\frac{N}{48\pi m}\int_x\left[
{\rm tr}\left\{\partial_{[\mu}\tilde{A}_{\nu]}
[\tilde{A}^\mu, \tilde{A}^\nu]\right\}
+O(\partial \frac{\partial^2}{m^2})\right].
\end{equation}

\subsection{Four-point gauge vertices}

\begin{figure}
\centerline{\hbox{\psfig{file=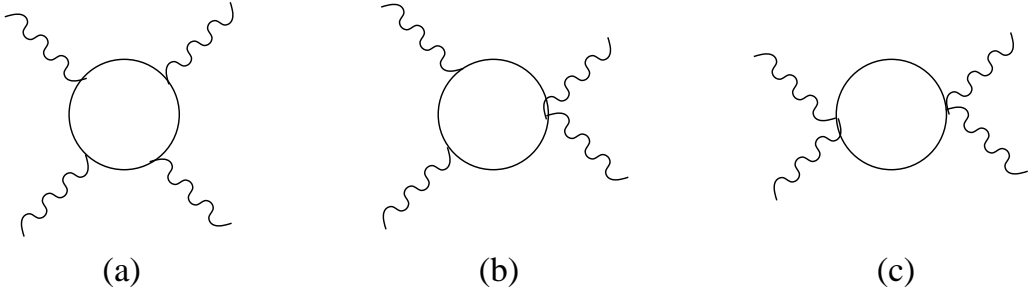,height=4.2cm}}}
\caption{Four-point gauge vertices. (a) $n=4$. (b) $n=3$. (c) $n=2$.}
\label{fig5}\vspace{12pt}
\end{figure}

The four-point gauge diagrams which contribute to the Yang-Mills action 
are shown in Fig.~\ref{fig5}. They are given by the following integrals:
\begin{eqnarray}
\mbox{(5a)} &=& -iN\frac{1}{4}{\rm Tr}\left[
G\tilde{A}_\mu \hat{\partial}^\mu G\tilde{A}_\nu \hat{\partial}^\nu 
G\tilde{A}_\rho \hat{\partial}^\rho G\tilde{A}_\sigma \hat{\partial}^\sigma
\right], \label{5a}\\
\mbox{(5b)} &=& -iN{\rm Tr}\left[G\tilde{A}_\mu \hat{\partial}^\mu 
G\tilde{A}_\nu \hat{\partial}^\nu 
G\tilde{A}_\rho\tilde{A}^\rho \right], \label{5b}\\
\mbox{(5c)} &=& -iN\frac{1}{2}{\rm Tr}
\left[G\tilde{A}_\mu\tilde{A}^\mu G\tilde{A}_\nu\tilde{A}^\nu\right].
\label{5c}
\end{eqnarray}
Calculation of the above integrals in the leading order of derivative 
expansion yields (See Appendix B) 
\begin{eqnarray}
\mbox{(5a)}+\mbox{(5b)}+\mbox{(5c)} &=& 
\frac{N}{4} f_T^{12}(0) \int_x~\Biggl[ 
-\frac{1}{2}W_{[\mu}C_{\nu ]}^* W^{[\mu}C^{\nu ]}
-b_1 W_{\mu}C^{*\mu}W_{\nu}C^{\nu}
\nonumber \\
&&+\frac{1}{4} C^\mu C_\mu C^{*\nu} C_\nu^*
-\frac{1}{2} \Bigl(2 -b_1 -\frac{3}{4}b_3 \Bigr)
C^\mu C_\mu^* C^\nu C_\nu^*
\Biggr]. \label{4point}
\end{eqnarray}
where we have defined $b_3$ as
\begin{equation}
b_3 \equiv \frac{f_T^{11}(0)+f_T^{22}(0)}{f_T^{12}(0)}.
\end{equation}
At the self-dual limit, Eq.\ (\ref{4point}) is simplified to 
be the following form
\begin{equation}
\mbox{(5a)}+\mbox{(5b)}+\mbox{(5c)}=\frac{N}{96\pi m}\int_x\left[
{\rm tr}\left\{[\tilde{A}_\mu, \tilde{A}_\nu]
[\tilde{A}^\mu, \tilde{A}^\nu]\right\}
+O(\frac{\partial^2}{m^2})\right].
\end{equation}

\subsection{Large-N effective action and the equations of motion}

Combining Eqs.\ (\ref{2point}), (\ref{3point}) and (\ref{4point}), 
we obtain the $U(1)_A \times U(1)_B$ gauge invariant effective action
\begin{eqnarray}
{\cal L}_{\rm eff} &=& 
-\frac{1}{4{g_A}^2}F_{\mu\nu}F^{\mu\nu}(A)
-\frac{1}{4{g_B}^2}F_{\mu\nu}F^{\mu\nu}(B)
+M_C^2 {V_\mu^*}V^\mu 
-\frac{1}{2}({\cal D}_{[\mu} V_{\nu]})^*({\cal D}^{[\mu} V^{\nu]})  
\nonumber \\
&&-\frac{2}{3}(b_1 +b_2)({\cal D}_\mu V_\nu)^*({\cal D}^\nu V^\mu)
-\frac{1}{3}(b_1 -2b_2)({\cal D}_\mu V^\mu)^*({\cal D}_\nu V^\nu)
\nonumber \\
&&-\Bigl(1-\frac{b_1}{3}-\frac{b_2}{3}\Bigr)
iV^\nu V^{*\mu}F_{\mu\nu}(A)             
+\Bigl(1-\frac{b_1}{3}+\frac{b_2}{3}\Bigr)
iV^{\nu} V^{*\mu}F_{\mu\nu}(B) \nonumber \\
&&+\kappa\, V^\mu V_\mu V^{*\nu}V^*_\nu
-2 \kappa\, \Bigl(2-b_1-\frac{3}{4}b_3 \Bigr)
V^\mu V_\mu^* V^\nu V_\nu^*,
\label{effgauge}
\end{eqnarray}
where the gauge couplings $g_A,g_B$ and the four-point coupling $\kappa$ 
are given by
\begin{equation}
g_A^2 \equiv \frac{1}{Nf_T^{11}(0)},\quad
g_B^2 \equiv \frac{1}{Nf_T^{22}(0)},\quad
\kappa \equiv \frac{1}{Nf_T^{12}(0)},
\end{equation}
and ${\cal D}_\mu$ denotes the $U(1)_A \times U(1)_B$ covariant 
derivative ${\cal D}_\mu \equiv\partial_\mu -iW_\mu$.
The $V$ field comes from the rescaling 
$\sqrt{Nf_T^{12}(0)}\,C_\mu \longrightarrow 2V_\mu$. 

The field equations derived from the above Lagrangian are given as follows;
\begin{eqnarray}
&& \partial^\mu F_{\mu\nu}(A) = g_A^2 J^+_\nu, \\
&& \partial^\mu F_{\mu\nu}(B) = g_B^2 J^-_\nu, \\
&& \partial^\mu F_{\mu\nu}(V) +b_1 \partial_\nu\partial_\mu V^\mu 
+M_C^2 V_\nu = \tilde{J}_\nu,
\end{eqnarray}
where the $U(1)_A \times U(1)_B$ current $J^\pm_\nu$ and the source 
current for the $V$ field $\tilde{J}_\nu$ are given such that
\begin{eqnarray}
J^+_\nu &=& i\Bigl( 1-b_1 -b_2 \Bigr)
\partial^\mu ( V_{[\mu} V^*_{\nu]} ) +ib_1 \Bigl[ V^*_\nu \partial_\mu 
V^\mu -V_\nu \partial_\mu V^{*\mu} \Bigr] \nonumber \\ &&
+i\Bigl[V^{\mu} F_{\mu\nu}(V^*) - V^{*\mu} F_{\mu\nu}(V) \Bigr]
+(b_1 -1) W^\mu  V_{(\nu} V^*_{\mu)}+2W_\nu V^*_\mu V^\mu,\\
J^-_\nu &=&  -J^+_\nu -\frac{2b_2}{3}~i\partial^\mu 
( V_{[\mu} V^*_{\nu]} ),\\
\tilde{J}_\nu &=& iW^\mu F_{\mu\nu}(V)
+i{\cal D}^\mu ( W_{[\mu} V_{\nu]} )
+ib_1 \Bigl[ \partial_\nu ( W_\mu V^\mu )
+W_\nu {\cal D}_\mu V^\mu \Bigr] \nonumber \\ &&
-i\Bigl(1-b_1 -b_2 \Bigr) V^\mu F_{\mu\nu}(A)
+i\Bigl(1-b_1 -\frac{b_2}{3} \Bigr) V^\mu F_{\mu\nu}(B) \nonumber \\ &&
+4\kappa \Bigl(2-b_1 -\frac{3}{4}b_3 \Bigr)V^\mu V^*_\mu V_\nu 
-2\kappa V^\mu V_\mu V^*_\nu.
\end{eqnarray}
The first two field equations require the current conservation 
$\partial_\nu J^{\pm\nu}=0$ which we can confirm by using all the field 
equations together with the identity
\begin{equation}
{\rm Im} \Bigl[ V^*_\nu \tilde{J}^\nu \Bigr]=
\frac{1}{2}\partial^\nu \Biggl[(b_1 -1)W^\mu V_{(\nu} V^*_{\mu)}
+2W_\nu V^*_\mu V^\mu \Biggr].
\end{equation}
Taking divergence of the third field equation yields 
\begin{equation}
(b_1 \Box +M_C^2)\,\partial_\mu V^\mu =\partial_\mu \tilde{J}^\mu.
\label{tach}
\end{equation}

Note that the parameter $b_1$ is negative (See Eqs.\ (\ref{ft12}) and 
(\ref{fl12}) in Appendix A). 
Therefore, if we turn off all the interactions 
($\partial_\mu \tilde{J}^\mu =0$), Eq.\ (\ref{tach}) tells us that the scalar 
mode of $V$ boson becomes a tachyon. As we will show shortly, the $V$ boson 
turns into the off-diagonal components of the enhanced $U(2)$ gauge bosons 
at the self-dual limit. In order to quantize the effective gauge theory 
(\ref{effgauge}), we have to take all the interaction terms into account 
even away from the self-dual points. In fact, $\partial_\mu \tilde{J}^\mu$ 
contains a term such as $A_\nu \partial^\nu (\partial_\mu V^\mu)$ 
which may possibly change the tachyonic behavior of the scalar mode.

\subsection{Yang-Mills action of the enhanced gauge symmetry}

At the self-dual limit $r = 1$, the effective action (\ref{effgauge}) 
turns into the $U(2)$ Yang-Mills action
\begin{eqnarray}
{\cal L}_{\rm eff}=
\frac{N}{96\pi m}\,{\rm tr}\,F_{\mu\nu}F^{\mu\nu}(\tilde{A}),
\end{eqnarray}
where 
$F_{\mu\nu}(\tilde{A})\equiv \partial_{[\mu}\tilde{A}_{\nu]}+
[\tilde{A}_\mu, \tilde{A}_\nu]$ is the field strength of the enhanced 
nonabelian $U(2)$ gauge symmetry. 
Away from the self-dual points $m_1\neq m_2$, the effective gauge 
action (\ref{effgauge}) is no longer written as a single trace of 
$U(2)$ matrix. 
However the three-point and four-point gauge interactions still 
preserve the $U(1)_A \times U(1)_B$ gauge invariance. 

We conclude this section by observing that the large-$N$ effective action 
is renormalizable in fewer than 3+1 dimensions. The only UV 
divergence is the one which arises in the gap equation and the other 
possible UV divergences in the vacuum polarization function are either 
forbidden by the gauge symmetry or related to the order parameters 
$\vec{v}_1$ or $\vec{v}_2$. The renormalization conditions (\ref{rg1}) 
and (\ref{rg2}) are enough to realize the UV finite large-$N$ theory. 
The higher order corrections in $1/N$-expansion can be systematically 
renormalized by using the counter terms which the large-$N$ effective 
action (\ref{eff1}) suffices. 
Unfortunately, in 3+1 dimensions, there arises a logarithmic divergence 
in the large-$N$ gap equations (See Eq.\ (\ref{4Dgap}) in Appendix A). 
This UV divergence prevents us from taking the continuum limit. To improve 
this involves modifying renormalization group equations by adding extra 
counter terms which absorb the logarithmic divergence and imposing a 
matching condition which requires the compositeness of dynamical gauge 
bosons \cite{tani,wein}.

\section{Conclusion and Discussion}
\setcounter{equation}{0}

We have performed the large $N$ path integral of a coupled $CP(N)$ model
with dual symmetry and analyzed the vacuum structure and renormalization
in the large-$N$ limit in 2+1 dimensions. The large-$N$ gap equation analysis
yields a solution with two decoupled gap equations. Consequently, we have the
dimensionless coupling constant $(u_1, u_2)$-plane separated into the four
regions with two UV fixed lines. Then we find the breaking patterns of the 
global $SU(N)$ and the local $H$ symmetries which are summarized in 
Table \ref{pattern}. 

Every transition between two of the four phases is 
the second order phase transition associated with the dynamical Higgs 
mechanism. However, the massive gauge boson which acquires a mass term 
through the Higgs mechanism is actually no longer stable and is dissociated 
into a pair of Nambu-Goldstone bosons (for example in the phase II, a massive 
$B$ boson decays into a pair of $\psi_2^\dagger$ and $\psi_2$).
Note that the origin of $C$ boson mass is not the Higgs mechanism but 
rather the explicit breaking parameter $r$, the radius (or inverse radius) 
of $CP(N)$. The $C$ boson is therefore a propagating massive vector field 
even in broken phases. 

We also have computed the effective gauge Lagrangian in the unbroken phase 
IV explicitly. The effective Lagrangian (\ref{effgauge}) tells us that other 
than dynamically generated gauge bosons $A$ and $B$, we have a propagating 
$C$ boson which acquires radiatively induced finite mass away from the UV 
fixed point. Besides, the RG analysis of Section 4 have shown that all the 
RG trajectories inside the phase IV flow into the self-dual UV fixed point 
where the two UV fixed lines intersect. Therefore we conclude that even 
if we start from the theory with two different $CP(N)$ radii, the theory 
favors the conformal fixed point with two coincident radii and 
the $U(1)_A \times U(1)_B$ gauge symmetry is enhanced to be a nonabelian 
$U(2)$ symmetry in UV limit. Note that the classical dual 
symmetry is not broken by the nonperturbative radiative corrections and 
survives in the effective action (\ref{effgauge}). 

The dynamical generation of the $C$ boson mass considered in this paper
is purely due to the finite radiative corrections, whereas
the conventional dynamical Higgs mechanism is known to be unsatisfactory due 
to the hierarchy problem. Therefore, our results could have some realistic 
applications, if the present analysis could be extended to 3+1 dimensions
\cite{tani,wein}. In this respect, it is useful to recall that one of the 
original motivations for the dynamical generation of gauge bosons through the 
nonlinear sigma model was to account for the gauge group which is large enough 
to accommodate the known standard model in ${\cal N}=8$ extended supergravity 
theory \cite{crem}. 
However, this theory has, although large enough, a non-compact 
sigma model sector and progress along this direction has been hampered by the
no-go theorem \cite{davis} which states that the dynamical generation
of gauge bosons does not occur for the non-compact target spaces. Therefore, 
it remains to be a challenging problem to overcome \cite{holt} the no-go 
theorem and extend our results to non-compact nonlinear sigma model in 3+1 
dimensions. 

\begin{table}
\begin{center}
\begin{tabular}{r|l|l}
\mbox{phase} & $[SU(N)]_{\rm global}$ & $[H]_{\rm local}$ \\ 
\hline 
I. & $SU(N-2)\times U(1)\times U(1)$ & fully broken \\
II. & $SU(N-1)\times U(1)$ & $U(1)_A$ \\
III. & $SU(N-1)\times U(1)$ & $U(1)_B$ \\
IV. & $SU(N)$ unbroken & $U(1)_A \times U(1)_B$ unbroken \\
\end{tabular}\vspace{12pt}
\caption{The breaking patterns of the global $SU(N)$ and the local $H$ 
symmetries. \label{pattern}}
\end{center}
\end{table}

\vspace{5mm}


T.I. was supported by the grant of Post-Doc.~Program, Kyungpook 
National University (2000). 
P.O. was supported by the Korea Research Foundation through project number 
DP0087.


\appendix

\renewcommand{\thesection}{Appendix \Alph{section}}
\renewcommand{\theequation}{\Alph{section}\mbox{.}\arabic{equation}}

\section{Dimensional regularization}

Throughout the calculation of vacuum polarization function and three- and 
four-point functions, we have used dimensional regularization which is 
simply calculating Feynman integrals in the space-time dimensionality 
$2\le D\le 4$. Two-dimensional results are obtained by introducing a small 
parameter $\epsilon \equiv (D-2)/2$ and taking the limit 
$\epsilon \to 0$. If we use another small parameter 
$\tilde{\epsilon} \equiv (4-D)/2$ and take the limit $\tilde{\epsilon}\to 0$, 
we can see four-dimensional results also. 

The vacuum polarization function in $D$ dimensions is given by the same 
Feynman integral (\ref{vacpol}), except that the momentum integration is 
now $D$-dimensional, and is calculated such that
\begin{equation}
\Pi_{\mu\nu}^{ij}(p)=\left(g_{\mu\nu}-\frac{p_\mu p_\nu}{p^2}\right)
\Pi_T^{ij}(p)+\left(\frac{p_\mu p_\nu}{p^2}\right)\Pi_L^{ij}(p),
\label{Dvacpol}
\end{equation}
with the transverse and longitudinal functions $\Pi_T$, $\Pi_L$ which are
obtained in $D$ dimensions as
\begin{eqnarray}
\Pi_T^{ij}(p) &=& 
\eta_D \Biggl[\frac{4}{D-2}\Biggr]
\Biggl[\frac{\,m_i^{D-2}+m_j^{D-2}}{2}
-\int_0^1 dx K^{\frac{D-2}{2}}\Biggr] \\ 
\Pi_L^{ij}(p)&=&
\eta_D \Biggl[\frac{\left(m_i^2 -m_j^2\right)^2}{p^2}\Biggr]
\Biggl[\frac{2}{D-2}\frac{\,m_i^{D-2}-m_j^{D-2}}{m_i^2 -m_j^2}
-\int_0^1 dx K^{\frac{D-4}{2}}\Biggr]
\end{eqnarray}
where $K \equiv x m_i^2 +(1-x)m_j^2 -x(1-x)p^2$ and 
$\eta_D \equiv \Gamma(2-{D\over2})/(4\pi)^{D \over 2}$.
Actually, the vacuum polarization function $\Pi^{ij}_{\mu\nu}$ includes 
an extra constant term
\begin{equation}
g_{\mu\nu} \eta_D \Biggr[\frac{2}{2-D}\Biggr] 
\left(m_i^{D-2}-m_j^{D-2}\right)
\end{equation}
which is asymmetric under interchanging $i$ and  $j$. However, this term  
completely vanishes in the effective action due to the cancellation 
between two off-diagonal terms, say $C^\mu \Pi^{12}_{\mu\nu} C^{*\nu}$ 
and $C^{*\mu} \Pi^{21}_{\mu\nu} C^{\nu}$, so that we ignored it in 
Eq.\ (\ref{Dvacpol}). 

The transverse and longitudinal functions are rewritten as
\begin{eqnarray}
\Pi^{ij}_{T}(p)=c_T^{ij} +p^2 f_T^{ij}(p),\quad
\Pi^{ij}_{L}(p)=c_L^{ij} +p^2 f_L^{ij}(p).
\end{eqnarray}
of which lowest order coefficients in momentum expansion are given by the 
integrals:
\begin{eqnarray}
c_T^{ij} &=& \eta_D \Biggl[\frac{4}{D-2}\Biggr]
\Biggl[\frac{\,m_i^{D-2}+m_j^{D-2}}{2}
-\int_0^1 dx M^{\frac{D-2}{2}}\Biggr], \\ 
c_L^{ij} &=&  \eta_D \Biggl[\frac{D-4}{2}\Biggr] 
\left(m_i^2 -m_j^2\right)^2 \int_0^1 dx\, x(1-x) M^{\frac{D-6}{2}}, \\
f_T^{ij}(0) &=& 2 \eta_D \int_0^1 dx\, x(1-x) M^{\frac{D-4}{2}}, 
\label{ft12}\\
f_L^{ij}(0) &=& -\eta_D \Biggl[\frac{(D-4)(D-6)}{8}\Biggr] 
\left(m_i^2 -m_j^2\right)^2 \int_0^1 dx\, x^2 (1-x)^2 M^{\frac{D-8}{2}},
\label{fl12}
\end{eqnarray}
where $M \equiv x m_i^2 +(1-x) m_j^2$. We find that all coefficients 
are symmetric under interchanging $i$ and $j$. Moreover, $c_T^{ij}$ and 
$c_L^{ij}$ are equal to each other ($c_T^{ij}=c_L^{ij} \equiv c^{ij}$) 
and vanish for $i=j$. Note also that $f_T^{12}(0)$ is always positive, 
while both $c^{12}$ and $f_L^{12}(0)$ are negative in $D<4$. 
The non-zero coefficients are calculated and 
determined as follows.\\

\noindent
Diagonal elements:
\begin{equation}
f_T^{11}(0)=\frac{\eta_D}{3}m_1^{D-4},\quad
f_T^{22}(0)=\frac{\eta_D}{3}m_2^{D-4}.
\end{equation}

\noindent
Non-diagonal elements:
\begin{eqnarray}
c^{12} &=& 2\eta_D 
\Biggl[\frac{1}{m_1^2 -m_2^2}\Biggr]
\Biggl[\frac{(D-4)(m_1^D -m_2^D)}{D(D-2)}
-\frac{m_1^{D-2}m_2^2-m_1^2 m_2^{D-2}}{D-2}\Biggr], \\
f_T^{12}(0) &=& 8\eta_D 
\Biggl[\frac{1}{m_1^2 -m_2^2}\Biggr]^3
\Biggl[\frac{m_1^{D+2} -m_2^{D+2}}{D(D+2)}
-\frac{m_1^D m_2^2-m_1^2 m_2^D}{D(D-2)}\Biggr], \\
f_L^{12}(0) &=& 2\eta_D 
\Biggl[\frac{1}{m_1^2 -m_2^2}\Biggr]^3
\Biggl[\frac{m_1^4 m_2^{D-2} -m_1^{D-2} m_2^4}{D-2}
+2\frac{(D-6)(m_1^D m_2^2-m_1^2 m_2^D)}{D(D-2)} \nonumber \\ &&
-\frac{(D-4)(D-6)(m_1^{D+2}-m_2^{D+2})}{D(D-2)(D+2)}
\Biggr].
\end{eqnarray}
Specifically, they are given for $D=2,3,4$ as follows.\\

\noindent
$D=2$:
\begin{eqnarray}
c^{12} &=& \frac{1}{2\pi}\Biggl[1-\frac{m_1^2+m_2^2}{m_1^2-m_2^2}
\ln \frac{m_1}{m_2}\Biggr],\\
f_T^{11}(0) &=& \frac{1}{12\pi m_1^2},\quad
f_T^{22}(0)~~=~~\frac{1}{12\pi m_2^2},\\
f_T^{12}(0) &=& \frac{1}{4\pi}\Biggl[\frac{1}{m_1^2 -m_2^2}\Biggr]^3
\Biggl[m_1^4-m_2^4-4m_1^2 m_2^2 \ln \frac{m_1}{m_2}\Biggr],\\
f_L^{12}(0) &=& \frac{3}{4\pi}\Biggl[\frac{1}{m_1^2 -m_2^2}\Biggr]^3
\Biggl[m_1^4-m_2^4-\frac{2}{3}\left(m_1^4+m_2^4+4m_1^2 m_2^2\right)
\ln \frac{m_1}{m_2}\Biggr].
\end{eqnarray}

\noindent
$D=3$:
\begin{eqnarray}
c^{12} &=& -\frac{1}{12\pi}\frac{(m_1-m_2)^2}{m_1+m_2},\\
f_T^{11}(0) &=& \frac{1}{24\pi m_1},\quad
f_T^{22}(0)~~=~~\frac{1}{24\pi m_2},\\
f_T^{12}(0) &=& \frac{1}{15\pi}\frac{m_1^2+3m_1 m_2+m_2^2}{(m_1+m_2)^3},\\
f_L^{12}(0) &=& -\frac{1}{20\pi}\frac{(m_1-m_2)^2}{(m_1+m_2)^3}.
\end{eqnarray}

\noindent
$D=4$:
\begin{eqnarray}
c^{12} &=& -\frac{1}{32\pi^2}\Biggl[m_1^2+m_2^2
-\frac{4m_1^2 m_2^2}{m_1^2-m_2^2} \ln\frac{m_1}{m_2}\Biggr],\\
f_T^{11}(0) &=& \frac{1}{48\pi^2}\Biggl[
\frac{1}{\tilde{\epsilon}}-\gamma-\ln 4\pi -\ln\frac{m_1^2}{\mu^2}
\Biggr],\\
f_T^{22}(0) &=& \frac{1}{48\pi^2}\Biggl[
\frac{1}{\tilde{\epsilon}}-\gamma-\ln 4\pi -\ln\frac{m_2^2}{\mu^2}
\Biggr],\\
f_T^{12}(0) &=& \frac{1}{48\pi^2}\Biggl[
\frac{1}{\tilde{\epsilon}}-\gamma-\ln 4\pi -\ln\frac{m_1 m_2}{\mu^2}
\Biggr] \nonumber\\ &&
+\frac{1}{48\pi^2}\Biggl[\frac{1}{m_1^2-m_2^2}\Biggr]^3 
\Biggl[\frac{1}{6}\left( 5m_1^4+5m_2^4-22m_1^2 m_2^2 \right)
\left( m_1^2-m_2^2 \right) \nonumber\\ &&
-\left( m_1^2+m_2^2 \right)\left( 
m_1^4+m_2^4-4m_1^2 m_2^2 \right)\ln \frac{m_1}{m_2}\Biggr],\\
f_L^{12}(0) &=& -\frac{1}{96\pi^2}\Biggl[\frac{1}{m_1^2-m_2^2}\Biggr]^3 
\Biggl[\left( m_1^2-m_2^2 \right)
\left( m_1^4+10m_1^2 m_2^2+m_2^4 \right) \nonumber\\ &&
-12 m_1^2 m_2^2 \left( m_1^2+m_2^2 \right)\ln\frac{m_1}{m_2} \Biggr].
\end{eqnarray}

In four dimensions there arises the same logarithmic divergence 
$\tilde{\epsilon}^{-1}$ in $f_T^{11}$, $f_T^{22}$ and $f_T^{12}$, which 
correspond to $U(1)_A$, $U(1)_B$ gauge couplings and the four-point 
self-coupling of $V$ boson, respectively. 
The same UV divergence also arises in the large-$N$ gap equation and 
breaks the renormalizability in $1/N$ expansion. 
Let us briefly look at how this goes on below. 
The dynamically generated boson masses $m_1$, $m_2$ are given by solving 
the gap equations in Section 3 in $D=4$ with setting $m_3 =0$. 
In the symmetric phase, we obtain
\begin{equation}
\frac{1}{Ng_i^2}=\frac{1}{16\pi^2}
\Biggl[ \Lambda^2 -m_i^2 \ln \frac{\Lambda^2}{m_i^2} \Biggr].
\label{4Dgap}
\end{equation}
The logarithmic divergence in the right hand side prevents us from 
taking the continuum limit where each of $m_i$ becomes independent of 
the cutoff $\Lambda$. This logarithmic divergence is the same as the one 
in the vacuum polarization. They are related to each other through the 
correspondence 
\begin{equation}
\ln \frac{\Lambda^2}{\mu^2} \leftrightarrow 
\frac{1}{\tilde{\epsilon}}-\gamma-\ln 4\pi
\end{equation}
between two regularization schemes.

\section{Three- and four- point vertices in the large-N limit}
\setcounter{equation}{0}

Three-point functions given in Eqs.\ (\ref{4a}) and (\ref{4b}) are 
combined into the following single integral in the leading order of 
momentum expansion.
\begin{equation}
N\sum_{ijk}\int_x \tilde{A}^\nu_{jk}(x)\tilde{A}^\rho_{ki}(x)
L^{ijk}_{\mu\nu\rho}(i\partial_x)\tilde{A}^\mu_{ij}(x).
\end{equation}
Each component of the integration kernel $L^{ijk}_{\mu\nu\rho}$ is 
determined such that
\begin{eqnarray}
L_{\mu\nu\rho}^{111}(p) &=& -\frac{4A}{3D}p_{[\nu}g_{\rho]\mu}, \\
L_{\mu\nu\rho}^{222}(p) &=& -\frac{4\bar{A}}{3D}p_{[\nu}g_{\rho]\mu}, \\
L_{\mu\nu\rho}^{112}(p) &=& -\frac{4I}{3D}p_{[\nu}g_{\rho]\mu}, \\
L_{\mu\nu\rho}^{221}(p) &=& -\frac{4\bar{I}}{3D}p_{[\nu}g_{\rho]\mu}, \\
L_{\mu\nu\rho}^{121}(p) &=& -\frac{4}{3D}\Biggl[
I_+ p_{\nu}g_{\rho\mu}-I_- p_{\rho}g_{\mu\nu}
+L^{\prime}p_{\mu}g_{\nu\rho}\Biggr], \\
L_{\mu\nu\rho}^{212}(p) &=& -\frac{4}{3D}\Biggl[
\bar{I}_+ p_{\nu}g_{\rho\mu}-\bar{I}_- p_{\rho}g_{\mu\nu}
+\bar{L}^{\prime}p_{\mu}g_{\nu\rho}\Biggr], \\
L_{\mu\nu\rho}^{211}(p) &=& -\frac{4}{3D}\Biggl[
I_- p_{\nu}g_{\rho\mu}-I_+ p_{\rho}g_{\mu\nu}
-L^{\prime}p_{\mu}g_{\nu\rho}\Biggr], \\
L_{\mu\nu\rho}^{122}(p) &=& -\frac{4}{3D}\Biggl[
\bar{I}_- p_{\nu}g_{\rho\mu}-\bar{I}_+ p_{\rho}g_{\mu\nu}
-\bar{L}^{\prime}p_{\mu}g_{\nu\rho}\Biggr],
\end{eqnarray}
where the coefficients are given by the following Feynman integrals
\begin{eqnarray}
A &=& \int^k k^2 [G_1 (k)]^3, \\
I_\pm &=& I \pm L,\quad L^{\prime}~~=~~L-\frac{3D}{4}Z,\\
I &=& \int^k k^2 [G_1 (k)]^2 G_2 (k), \\
L &=& \frac{4}{D+2} \int^k (k^2)^2 [G_1 (k)]^2 G_2 (k) 
\Bigl[G_1 (k)-G_2 (k)\Bigr], \\
Z &=& \int^k \Biggl[ G_1 (k) G_2 (k) -\frac{4}{D} 
k^2 G_1 (k) [G_2 (k)]^2 \Biggr].
\end{eqnarray}
where $\int^k \equiv \int d^D k/(2\pi)^D$ and 
$G_i (k) \equiv 1/(k^2 +m_i^2)$. The integrals with a bar symbol 
are obtained by switching $m_1$ and $m_2$, for example, 
$\bar{I}=I|_{G_1 \leftrightarrow G_2}$. We can also confirm that 
$\bar{I}_\pm \equiv I_\pm$ and $\bar{L}^\prime \equiv L^\prime$. 
Computing the above integrals provides the following matching equations 
which yields Eq. (\ref{3point}) in Section 5.2. 
\begin{eqnarray}
f_T^{12}(0) &=& -\frac{4i}{3D}I_-,\\
f_L^{12}(0) &=& \frac{4i}{3D}(I_+ -I_- +L^{\prime}).
\end{eqnarray}
Note that $A$ and $\bar{A}$ do not contribute to the effective action 
after contracting with gauge fields. The integrals $I$ and $\bar{I}$ 
provide non-minimal gauge interactions which cannot be written 
in terms of the covariant derivative ${\cal D}_\mu =\partial_\mu 
-iA_\mu +iB_\mu$ in Section 5.4.

Similarly, four-point functions given in Eqs.\ (\ref{5a}), (\ref{5b}) 
and (\ref{5c}) are cast into the following single integral in the 
leading order of momentum expansion.
\begin{equation}
N\sum_{ijk}\int_x \Biggl[
\tilde{A}^\mu_{ij}(x)\tilde{A}_{\mu kl}(x)
\tilde{A}^\nu_{jk}(x)\tilde{A}_{\nu li}(x) W^{ijkl}
+\tilde{A}^\mu_{ij}(x)\tilde{A}_{\mu jk}(x)
\tilde{A}^\nu_{kl}(x)\tilde{A}_{\nu li}(x) L^{ijkl}\Biggr].
\end{equation}
Each component of the integral kernels $W^{ijkl}$ and $L^{ijkl}$ 
is given by the following Feynman integrals
\begin{eqnarray}
L^{ijkl} &=& 2W^{ijkl}+W^{ijk}+W^{ik},\\
W^{ik} &=& -\frac{i}{2}\int^k G_i (k) G_k (k),\\
W^{ijk} &=& \frac{4i}{D}\int^k k^2 G_i (k) G_j (k) G_k (k),\\
W^{ijkl} &=& -\frac{4i}{D(D+2)}\int^k (k^2)^2 G_i (k) G_j (k) 
G_k (k) G_l (k).
\end{eqnarray}
Note that $W^{ik}$, $W^{ijk}$ and $W^{ijkl}$ are all completely symmetric 
tensors. Again computing the above integrals provides the following matching 
conditions which yields Eq. (\ref{4point}) in Section 5.3. 
\begin{eqnarray}
f_T^{12}(0) &=& 2W^{1122}~~=~~-L^{1112}-L^{1211},\\
f_L^{12}(0) &=& 2W^{1122}-4W^{1112}-L^{1121}-L^{2111} \nonumber\\
&=& L^{1122}+L^{2211}-L^{1112}-L^{1211}.
\end{eqnarray}
The four-point self-couplings of $V$ bosons, say 
$V^*_\mu V^{*\mu} V_\nu V^\nu$ and $V^*_\mu V^\mu V^*_\nu V^\nu$, 
which cannot be obtained from the covariant derivative ${\cal D}_\mu$, 
are given by $W^{1212}+W^{2121}$ and $L^{1212}+L^{2121}$ respectively.
 
\newpage



\begin{thebibliography}{99}

\bibitem{call} S. Coleman, J. Wess, and B. Zumino, Phys. Rev. B 177 (1969)
2237; C. Callan,  S. Coleman, J. Wess, and B. Zumino, Phys. Rev. B 177
(1969) 2247.

\bibitem{band} M. Bando, T. Kugo, and K. Yamawaki, Phys. Rep. 164 (1988) 217.

\bibitem{wilz} D. J. Gross and F. Wilczek, Phys. Rev. Lett. 30 (1973) 1343;
Phys. Rev. D 8 (1973) 3633; 
 H. D. Politzer, Phys. Rev. Lett. 30 (1973) 1346.


\bibitem{bela} A. A. Belavin, A. M. Polyakov, A. S. Schwartz, and Yu. S. 
Tyupkin,  Phys. Lett. B 59 (1975) 85; A. A. Belavin and A. M. Polyakov,
JETP Lett. 22 (1975) 245.


\bibitem{cole} S. Coleman,  Aspects of Symmetry
(Cambridge Univ. Press, Cambridge, 1985).

\bibitem{dada} A. D'Adda, M. L\"uscher, and P. Di Vecchia, Nucl. Phys. B 146
(1978) 63; ibid 152 (1979) 125;
E. Witten, Nucl. Phys. B 149 (1979) 285.


\bibitem{gree} M. B. Green, J. H. Schwarz, and E. Witten, 
Superstring Theory Vol. I and II 
(Cambridge Univ. Press, Cambridge,1987);
J. Polchinski, String theory Vol. I and II
(Cambridge Univ. Press, Cambridge, 1998).

\bibitem{give} A. Giveon, M. Porrati, and E. Rabinovici,
Phys. Rep. 244 (1994) 77.

\bibitem{zakr} W. J. Zakrzewski, Low Dimensional Sigma Models
(IOP Publishing Ltd, Bristol, 1989).

\bibitem{ior} T. Itoh, P. Oh, and C. Ryou, Phys. Rev. D 64 (2001)
045005.


\bibitem{golo} H. Eichenherr, Nucl. Phys. B 146 (1978) 215; 
V.L. Golo and A.M. Perelomov, Phys. Lett. B 79 (1978) 112.

\bibitem{gava} A. J. Macfarlane, Phys. Lett. B 82 (1979) 239;
R. D. Pisarski, Phys. Rev. D 20 (1979) 3358 ;
E. Gava, R. Jengo, and C. Omero,
Nucl. Phys. B 158 (1979) 381;
E. Brezin, S. Hikami, and J. Zinn-Justin, Nucl. Phys. B 165 (1980) 528;
S. Duane, Nucl. Phys. B 168 (1980) 32; G. Duerksen, Phys. Rev. D 
 24 (1981) 926. 

\bibitem{rose} B. Rosenstein, B. Warr, and S.H. Park,
Phys. Rep. 205 (1991) 59.


\bibitem{aref} I. Ya. Aref'eva and S. I. Azakov, Nucl. Phys. B 162 (1980)
298; I. Ya. Aref'eva, Ann. Phys. (N.Y.) 117 (1979) 393.
I. Ya. Aref'eva, V.K. Krivoshchekov, P. B. Medvedev, 
Theor. Math. Phys.  40 (1980) 565.


\bibitem{itoh} T. Itoh and P. Oh, Phys. Lett. B 491 (2000) 362;
Phys. Rev. D 63 (2001) 025019, hep-th/0006163 .

\bibitem{oh1} P. Oh and Q.-H. Park, Phys. Lett. B 383 (1996) 333;
ibid B 400 (199) 157; (E) 416 (1998) 452;
P. Oh, J. Phys. A: Math. Gen. 31 (1998) L325;
Rep. Math. Phys. 43 (1999) 271; Phys. Lett. B 464 (1999) 19.


\bibitem{bal} The coadjoint orbit formulation
of nonlinear sigma model on $G/H$ proved to be very useful in many aspects of
the subject, and the hidden local symmetry
with $G_{global}\times H_{local}$ is explicitly built in the
coadjoint orbit variable. See E. Cremmer and B. Julia, 
Phys. Lett. B 80 (1978) 48; Nucl. Phys. B 159 (1979) 141;
A. P. Balachandran, A. Stern, and C. G. Trahern, 
Phys. Rev. D 19 (1979) 2416; 
M. Bando, T. Kugo, and K. Yamawaki, Prog. Theo. Phys. 73 (1985)
1541 for the hidden local symmetry approach.


\bibitem{helg} S. Helgason, Differential Geometry, Lie Groups, and
Symmetric spaces (Academic Press, 1978).

\bibitem{jack} See R. Jackiw, Int. J. Mod. Phys. B 14 (2000) 2011
and references therein;J.-M. Chung and P. Oh,
Phys.Rev. D 60 (1999) 067702; R. Jackiw and Alan Kostelecky,
Phys. Rev. Lett. 82 (1999) 3572; W. F. Chen, Phys. Rev. D 60 (1999) 085007;
M. Perez-Victoria, Phys. Rev. Lett. 83 (1999)  2518;
J.M. Chung, Phys. Lett. B 461 (1999) 138.  


\bibitem{tani} M. Bando, Y. Taniguchi, and S. Tanimura,
Prog. Theor. Phys. 97 (1997) 665.


\bibitem{wein} See also S. Weinberg, Phys. Rev. D 56 (1997) 2303; S. V. Ketov,
Nucl. Phys. B 544 (1999) 181.


\bibitem{crem} E. Cremmer and B. Julia, Ref. \cite{bal} 

\bibitem{davis} A. C. Davis, A. J. Macfarlane, and J. W. van Holten,
Phys. Lett. B 125 (1983) 151;
A. C. Davis, M. D. Freeman, and A. J. Macfarlane, Nucl. Phys. B 258
 (1985) 393.

\bibitem{holt} See J. W. van Holten, Phys. Lett. B 135 (1984) 427 and
Nucl. Phys. B 258 (1984) 307 for earlier attempts to overcome the no-go 
theorem.
\end{thebibliography}
\end{document}